\documentclass[11pt,a4paper,oneside]{scrartcl}

\usepackage{natbib}
\usepackage{graphicx}
\usepackage{amssymb,amsmath}

\usepackage{scrpage2}
\deftripstyle*{myheadings}{\sffamily\upshape\small Sch\"onhof, Treiber, Kesting, and Helbing}{}{}
{\sffamily\upshape\small http://www.trb.org}
{\sffamily\upshape\small Annual Meeting of the Transportation Research Board}{\sffamily\upshape\small January 2007}
\usepackage{caption2}

\setcaptionwidth{1.0\textwidth}

\newcommand{\avg}[1]{\langle #1 \rangle}

\author{Martin Sch\"onhof, Martin Treiber, Arne Kesting, and Dirk Helbing\\[1ex]Institute for Transport \& Economics\\
Technische Universit\"at Dresden\\
Andreas-Schubert-Strasse 23\\
D-01062 Dresden, Germany}

\title{Autonomous detection and anticipation of\\jam fronts from 
messages propagated by inter-vehicle communication}
\date{November 15, 2006}
\begin{document}

\maketitle

\abstract{In this paper, a minimalist, completely distributed freeway traffic
information system is introduced. It involves an autonomous,
vehicle-based jam front detection, the information transmission via
inter-vehicle communication, and the forecast of the spatial position
of jam fronts by reconstructing the spatiotemporal traffic situation
based on the transmitted information.  The whole system is simulated
with an integrated traffic simulator, that is based on a realistic
microscopic traffic model for longitudinal movements and lane
changes. The function of its communication module has been explicitly
validated by comparing the simulation results with analytical
calculations.  By means of simulations, we show that the algorithms
for a congestion-front recognition, message transmission, and
processing predict reliably the existence and position of jam fronts
for vehicle equipment rates as low as 3\%. A reliable mode of
operation already for small market penetrations is crucial for the
successful introduction of inter-vehicle communication. The short-term
prediction of jam fronts is not only useful for the driver, but is
essential for enhancing road safety and road capacity by intelligent
adaptive cruise control systems.}

\newpage

\section{\label{sec:intro}Introduction}
%
Inter-vehicle communication (IVC) is widely regarded as a promising
concept for transmitting traffic-related information. There are
essentially two types of future applications that inspire the research
on IVC: Applications in traffic safety such as automated reaction to
an emergency incident, cooperative, autonomous driving and platoon
formation on freeways rely on fast communication and information
transmission between single vehicles 
\citep{SmartCars-Varaiya,RaoVaraiya-1993,Rao-ACC,Aoki1996,Tank1997}.
On the other hand, applications towards advanced traveler information
systems, dynamic routing, or entertainment applications do not depend
critically on information transmission times.

Recently, another application field of IVC has been proposed in the
context of strategically operating adaptive cruise control (ACC)
systems that change their driving characteristics automatically on an
intermediate timescale according to the local traffic situation
\citep{ACC-Arne-TRB07}. While currently available ACC systems aim to
enhance the comfort and safety of driving, their impact on the
capacity and the stability of traffic flow on freeways has moved into
the focus of traffic research
\citep{TGF05-ACC,Davis-ACC,VanderWerf-TRR-2001,Treiber-aut,marsden-ACC}.
Receiving traffic-related messages via IVC could help ACC systems to
recognize relevant traffic situations faster and more reliably. This
allows ACC-equipped vehicles to drive with an 'intelligent' driving
strategy that adapts the ACC parameters to the current traffic
situation thereby changing the 'driving style'. For example, if the
positions of jam fronts were known in advance, an ACC system could
brake earlier and smoother when approaching the upstream jam front to
increase traffic safety. In contrast, it could keep smaller time gaps
to the leader when leaving the jam at the downstream jam front to
increase the jam outflow (discharge rate), while staying at a normal
operation characteristics in all other situations. Furthermore,
ACC-equipped cars acting as 'floating cars' are also able to detect
the position of jam fronts and, consequently, may spread such
information via IVC.  However, as IVC will start on the basis of a
small number of equipped vehicles, it is crucial to investigate the
functionality and the statistical properties of the message hopping
processes under such conditions. These questions have been
investigated recently within the German research project INVENT
\citep{INVENT}.

Fast and reliable information spreading is a necessary precondition
for a successful implementation of all mentioned IVC-based
applications. Assuming a sufficient market penetration, IVC offers the
possibility of a decentralized and robust traffic information system,
where the data is collected, evaluated, and distributed autonomously
by each single car. Note, that estimations of the necessary market
penetration for IVC depend on the application, that means on the type
of information that is transmitted and how this information is used.
The transmission of messages within a dynamic ad-hoc network of
vehicles has been investigated on different levels of abstraction with
respect to protocol design \citep{KimNaka1997,XuMak2004} and message
propagation efficiency. The latter aspect that is also addressed in
this contribution has been investigated by simulations and by
analytical calculations in different studies
\citep{briesemeister:iv:00,wisch,Wu2004IEEE,WuLee2005,JinRecker2005}.
We refer to \cite{Yang2005} for a short and thorough literature
overview.
 
In this contribution, we present an IVC-based application of
vehicle-based jam-front detection and prediction. By means of a
microscopic traffic simulation, we simulate the whole chain of
information generation, transmission, and interpretation based on a
small fraction of vehicles equipped with IVC. Single vehicles in the
traffic simulation detect jam fronts and generate traffic-related
messages based on their locally available floating-car data. These
messages are propagated further upstream via IVC mainly by cars of the
opposite driving direction. Finally, the received information is used
for a reconstruction and short-term prediction of the expected traffic
situation further downstream which can serve as basis for determining
the appropriate ACC strategy as discussed by \cite{ACC-Arne-TRB07}.
We will investigate the prediction error for an equipped car as a
function of the distance to the predicted jam front. Notice that this
individual short-term traffic forecast is of general interest for the
driver.

Our paper is structured as follows: In Sec.~\ref{sec:model}, we
introduce the model of IVC and the generation of traffic-related
messages based on floating-car data. In Sec.~\ref{sec:messageProp},
the properties of the simulated message transmissions are validated by
comparison to analytical calculations. We also discuss the IVC
parameters. In Sec.~\ref{sec:result}, the proposed algorithm for
detecting and predicting jam fronts is applied to a basic traffic
scenario. Finally, we conclude with a short discussion and an outlook.

\section{\label{sec:model}Microscopic modeling of inter-vehicle communication and message generation}

\subsection{\label{IVC}Characteristics of inter-vehicle communication and its implementation}

\begin{figure}
\begin{center}
  \includegraphics[width=11cm]{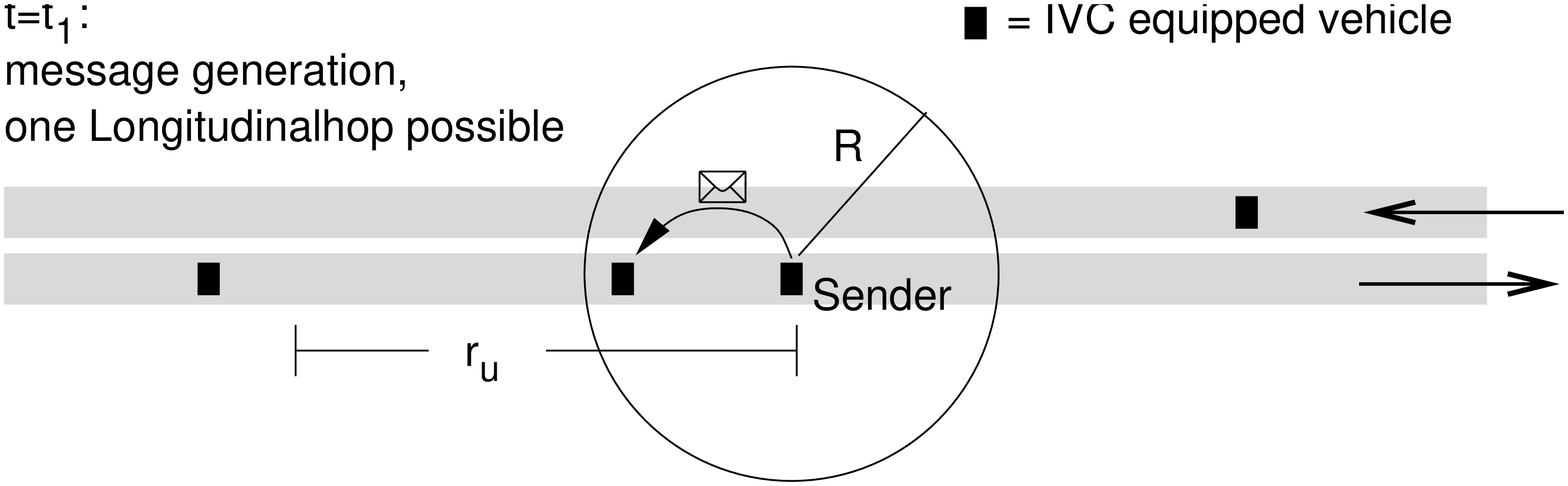}
  \includegraphics[width=11cm]{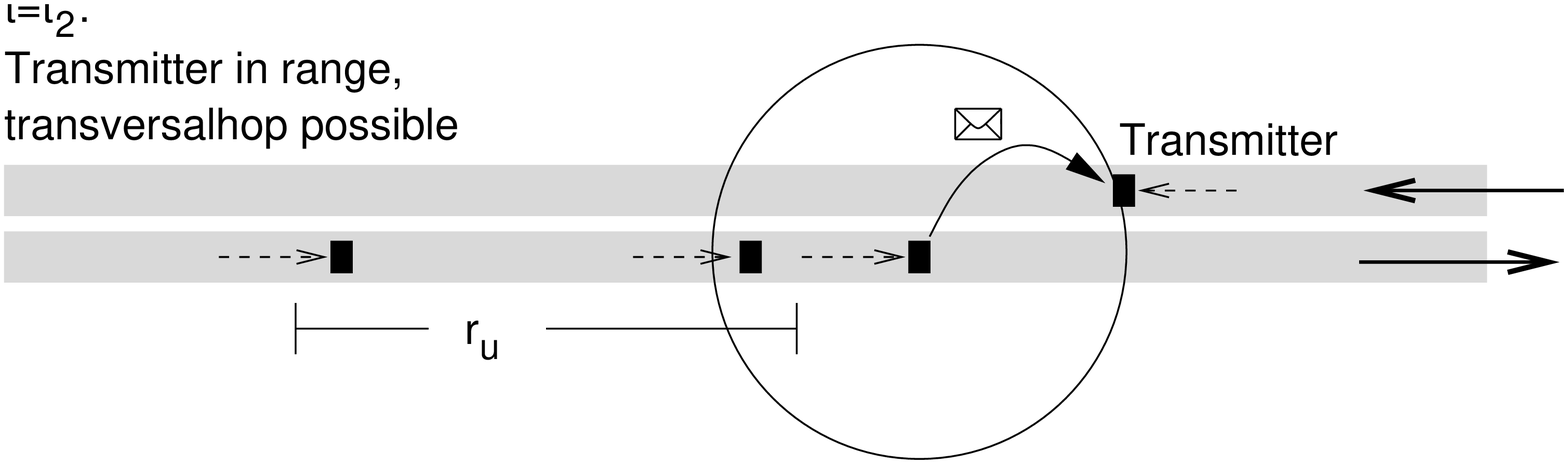}
  \includegraphics[width=11cm]{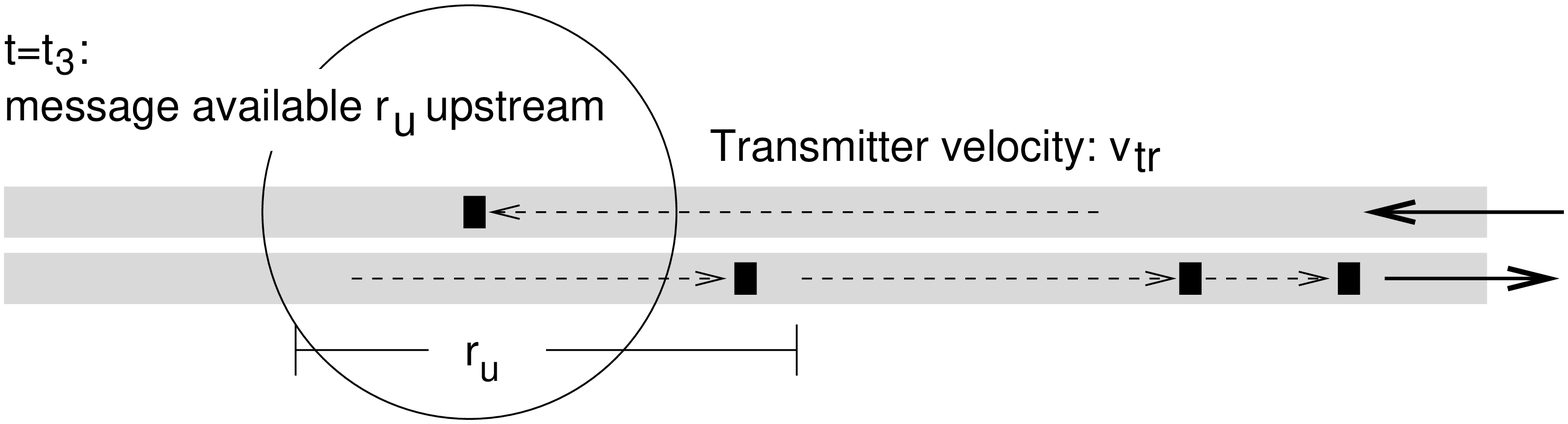}\\[1cm]
  \includegraphics[width=11cm]{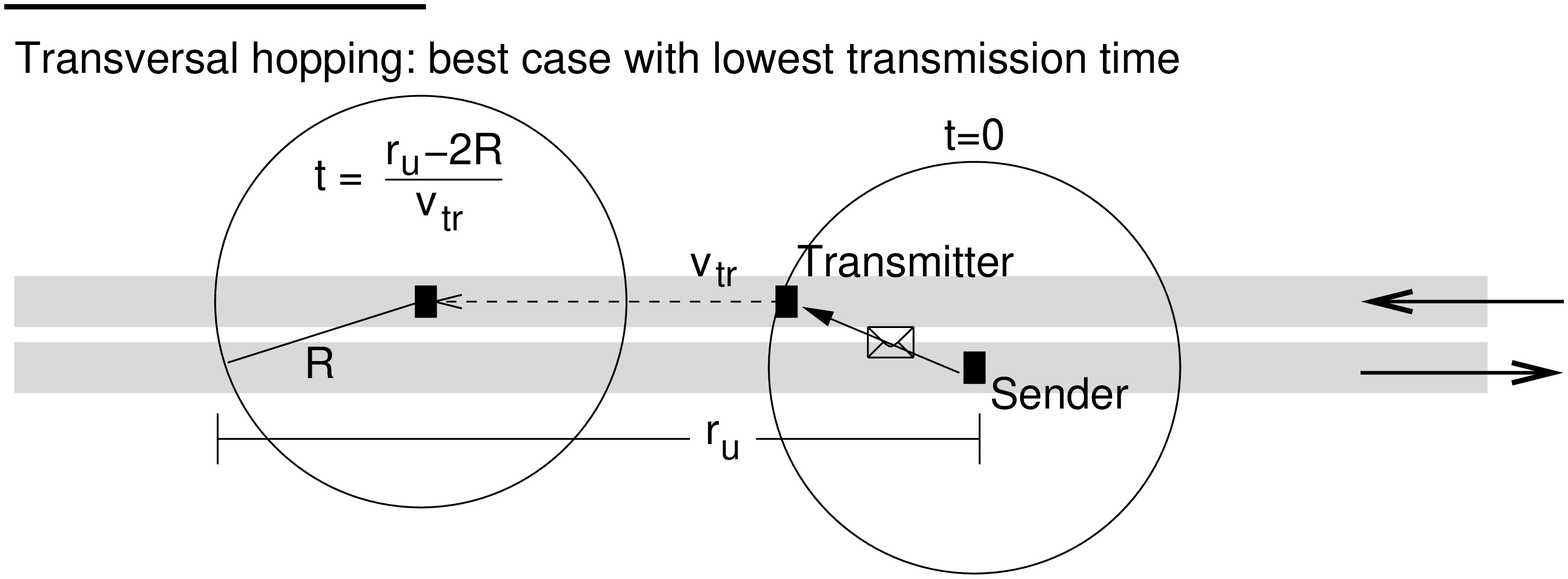}
\end{center} 
\caption{\label{fig:sketch}Basic mechanisms for the transport of a
  traffic-related message on a freeway: The sender (e.g., having
  recognized an upstream jam front) generates a message and starts
  broadcasting it. The message may be received by a subsequent car via
  a {\em longitudinal hop}.  At time $t_2$, the message is received by
  an equipped transmitter car via a {\em transversal hop}. The
  characteristic quantity of the communication process is the time
  $\tau=t_3-t_1$ from the message generation until it is available for
  the first time at a {\em user distance} $r_\text{u}$ upstream of the
  position where the message has been generated.  The minimal time $\tau$
  for transversal hopping is realized for a configuration as shown
  in the last sketch: The transmitter (velocity $v_\text{tr}$) must be
  at the optimal position at $t=0$, i.e., a distance $R$ upstream of
  the sender. The message is available, when the transmitter has
  covered the distance $r_{\rm u}-2R$, i.e., after a time
  $\tau_{\rm min}=(r_{\rm u}-2R)/v_{\rm tr}$. Notice that only IVC-equipped
  vehicles are depicted in the illustrations.}
\end{figure}

In this section, the model of inter-vehicle communication (IVC) is
introduced. In the context of freeway traffic, messages have to travel
upstream in order to be valuable for their receivers. In general,
there are two strategies, how a message can be transported upstream
via IVC as displayed in Fig.~\ref{fig:sketch}: Either the message hops
from one IVC car to a subsequent IVC car within the same driving
direction ({\it 'longitudinal hopping'}), or the message hops to an
IVC-equipped vehicle of the other driving direction, which takes the
message upstream and delivers it back to cars of the original driving
direction ({\it 'transversal hopping'}).

%
%

For a low density of equipped cars, i.e., for a small market
penetration, an instantaneous multiple longitudinal hopping is not
very likely due to gaps that are larger than a given limited broadcast
range \citep{Dousse2002}. It is argued, that the gaps in multi-lane
traffic may be bridged after a while because of different velocities
of the equipped cars. However, this is a minor effect: For low
densities of equipped vehicles, a message propagates in driving
direction just with the velocity not larger than the velocity of the
fastest cars \citep{Wu2004IEEE}. Therefore, we conclude that the
longitudinal hopping against the driving direction does not work for
low densities of equipped vehicles, because the mechanism for bridging
gaps is to slow for a real backward propagation of the messages -- the
normal downstream movements of messages inside the cars cannot be
outmatched by the wireless communication in upstream direction. Thus,
longitudinal hopping will only have some impact in combination with
transversal hopping in order to enhance the message propagation in the
traffic stream of the opposite driving direction.  However, in case of
dense traffic conditions (which is to be distinguished from a high
density of IVC-equipped vehicles), the variation in the velocities of
the vehicles is limited. As a consequence, even in the limit of long
timescales, the obtained propagation velocity is similar to the mean
velocity of the transmitter vehicles \citep{Wu2004IEEE}.

In the following, the used microscopic model for the IVC-related
processes is outlined: Every 2 seconds, messages are exchanged between
IVC-equipped vehicles within a limited broadcast range $R$. Each car
sends all its stored messages, and the default broadcast range is set
to $250\,$m. These assumptions are well justified with regard to the
current technological possibilities of data exchange between vehicles,
cf. the next Sec.~\ref{sec:realWorld}. Furthermore, we neglect the
width of the road in the calculations and simulations, i.e., the
broadcast range $R$ refers to the longitudinal distances.

Because of the low efficiency of longitudinal hopping, we restrict
ourselves to transversal hopping processes. Each car accepts only
messages from the other driving direction, either as a transmitter
vehicle (after the first transversal hop), or as a user that receives
information about its own driving direction (after a second
transversal hop).  All other messages will be discarded directly after
reception. Furthermore, messages related to events at an already
passed position, and messages that are older than 10 minutes are
deleted as well. As the routing in this system is obviously given by
the two traffic streams in opposite directions, no further rule is
necessary for modeling the message exchange process.

\subsection{\label{sec:realWorld}Technological basis for inter-vehicle communication}

The assumptions about the exchange of small traffic-related data
packages are justified by the experimentally proven possibilities of
data transmission between vehicles on the freeway within the IEEE
802.11b standard.  Operating WLAN equipment on a freeway with an
external antenna in a broadcast-like modus, i.e., using IP/UDP (user
datagram protocol), allows a data throughput of about 1Mb/s. This
holds for distances of about 300m, and even for cars moving in
different driving directions with a relative speed difference of 200
km/h \citep{Singh2002}. Despite high relative velocity differences
between two vehicles, the total transmission of more than three MB
data within one encounter has been reported \citep{OttKut2004},
although the available time within the broadcast range decreases with
the relative velocity, and some time is needed to associate to the
communication channel. Without giving quantitative values,
\cite{Yvonne2005} predict for a high vehicle density and for a high
market penetration rate a breakdown of of the communication due to too
many users using the available bandwidth. Such scenarios indeed have
not yet been investigated empirically.  The reduction of transmission
power, or simply sending messages more rarely may avoid the breakdown
of the communication channel: For the jam front prediction, it is not
necessary to receive a message of every single car, that has detected
the front. In addition, the new pending standard IEEE 802.11p, for
``Wireless Access for the Vehicular Environment'' comes with a much
more fast and efficient protocol than the IP protocol: The ``Dedicated
Short Range Communications'' (DSRC) is a concept specifically designed
for automotive use.  According to \cite{XuMak2004}, i.e., it is no
problem to transmit in normal freeway traffic (4 lanes, 33veh/km/lane)
small data packages of 400 Bytes from each car to each other car
inside a broadcast range of 150 m within a fraction of a second, and
with a loss rate of below 1\%.  Notice, that a message containing
traffic-related information, i.e., about a jam front, has a size of
the order of a few hundred Bytes rather than Kilobytes.

\subsection{\label{sec:fronts}Congestion fronts: Recognition and generation of traffic-related messages}

Since our focus is to use the traffic information as input for
traffic-adaptive ACC systems, the crucial
events to be detected and transmitted are the positions of jam fronts.

There are cases, when an (expected) jam front position may be
anticipated, e.g., if a car got stuck into a traffic jam upstream
bottleneck, that is well-known for causing congestion.  In many cases,
jam front positions can only be exactly detected by cars passing the
location.  Figure~\ref{fig:jam} illustrates three examples of
different congestion fronts:
%
\begin{figure}[th!]
\begin{center}
  \includegraphics[width=13cm]{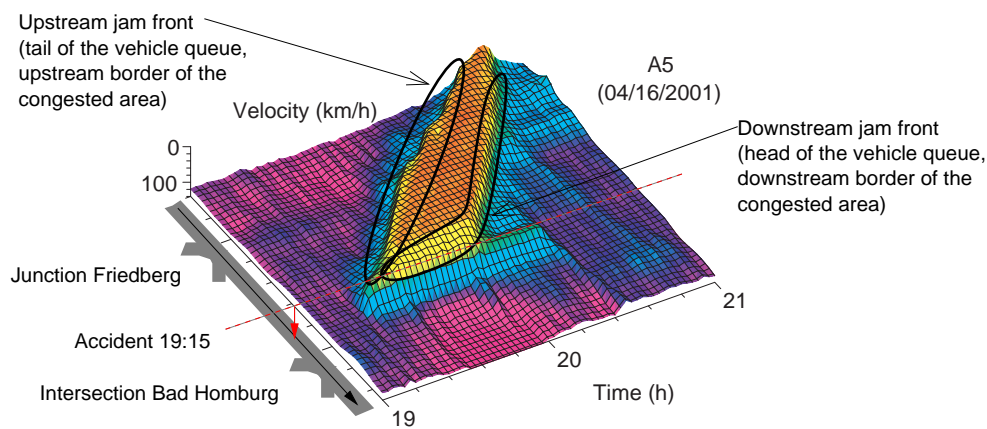}
\end{center}
\caption{\label{fig:jam}Example for the spatiotemporal dynamics of
  freeway jams for illustrating different types of jam fronts (see the
  main text). This traffic jam on the German freeway A5 between Kassel
  and Frankfurt in direction South was caused by a blockage of the
  most-right of altogether three lanes after an accident occurred as
  noted in the sketch of the freeway.  1-minute averaged date of the
  velocity had been recorded by double loop detectors with an average
  distance of 1 km. The shown velocity field has been obtained by
  using an adaptive smoothing method for interpolation of macroscopic
  traffic data between the detectors \citep{Treiber-smooth}. The
  driving direction of the cars is indicated by the black arrow.
  Notice the inverted vertical axis for matters of better
  illustration.}
\end{figure}
%
%
\begin{enumerate}
\item 
  A downstream jam front is pinned at some bottleneck,
  e.g., at the location of an incident referring to the straight line
  in Fig.~\ref{fig:jam}.
\item A downstream jam front (a dissolution front of congested
  traffic) is propagating against the traffic flow with a characteristic speed of about $-15\,$km/h
  \citep{Kerner-Rehb96,Phase,Martin-empStates}.
\item An upstream jam front is moving with an propagation speed that
 depends on the traffic flow upstream of the jam and the flow in the
 jam.
\end{enumerate}
Downstream jam fronts are normally straight lines in the
spatiotemporal plane. These fronts either are fixed at a bottleneck,
or move with a constant velocity of approximately -15 km/h -- apart
from few cases of a 'moving bottleneck' where the propagation velocity
can assume other values but is constant as well.  This fundamental
feature of traffic flow dynamics is found in most of the empirical
work and is also reflected in the traffic models from the very
beginning \citep{LiWhi55}. The velocity of downstream jam fronts can
even be constant for hours.  For stationary inflow conditions,
upstream jam fronts also have a constant velocity. It may vary between
between -25~km/h (the high value of -40 km/h reported by
\cite{BertMalik2004} has not been observed in other studies), and the
velocity of freely driving vehicles (if the inflow is vanishing and a
cluster of vehicles is accelerating). In the most cases, the
approximation of a constant front velocity is justified, because the
traffic demand normally does not change significantly on time scales
of 5--10 minutes.

For the detection and prediction of jam fronts, their spatial
extension has to be considered \citep{TSMunozDaganz-03}. Due to the
discrete nature of traffic, the jam 'front position' as a continuous
line in time and space can only be thought as an abstract result of an
averaging process based on vehicle trajectories.

In the following, a model for the jam front detection based on
floating-car data is presented: The jam front is characterized by the
time and the location, when a passing car starts to brake or to
accelerate. In order to reliably detect these acceleration and
deceleration processes, and to minimize the number of 'false alarms',
each car smoothes its floating-car data, i.e., its velocity $v(t)$
using an exponential moving average (EMA):
\begin{equation}\label{eq:EMA}
v_\text{EMA}(t) = \frac{1}{\tau} \int\limits^{t}_{-\infty} dt' e^{-(t-t')/\tau} \,v(t'),
\end{equation}
with a relaxation time $\tau=10$~s. The EMA allows for an efficient
real-time update by using an explicit integration scheme for the
corresponding ordinary differential equation
\begin{equation}
\frac{d}{dt} v_\text{EMA} = \frac{v-v_\text{EMA}}{\tau}.
\end{equation}

The detection of an upstream or downstream jam front relies on a {\it
change} in speed compared to the exponentially averaged past of the
speed. An upstream jam front is therefore given, when for the first time
\begin{equation}\label{eq:up}
v(t) - v_\text{EMA}(t) < - \Delta v_\text{up}
\end{equation}
holds, with $\Delta v_\text{up}=15$~km/h.
A downstream jam front is identified by the first notice of an acceleration
period,
\begin{equation}\label{eq:down}
v(t) - v_\text{EMA}(t) > \Delta v_\text{down}.
\end{equation}
with $\Delta v_\text{down} = 10\,$km/h. When a congestion front is
detected, a corresponding message containing position, time, and
jam-front type is generated. This message is repeatedly broadcasted
until it is discarded after 10 minutes.

\section{\label{sec:messageProp}Statistics of message propagation via the opposite driving direction}

In this section, we will compare simulation results for the efficiency
of IVC via transversal hopping with analytical results
\citep{IVC-Goldrain06}.  Moreover, we will shortly
discuss the parameters of the system.

When the proportion $\alpha$ of vehicles equipped with IVC is low, the
positions of the IVC-equipped cars can be assumed to be statistically
independent of each other. Therefore, the arrival process of an
equipped vehicle at a given cross-section is a Poisson process. This
holds even for high traffic densities, where the positions of
neighboring vehicles are highly correlated.  We define the {\em
density of equipped cars} $\lambda$ in one driving direction by
$\lambda=\rho\alpha$, where $\rho$ is the traffic density over all
lanes.  As a consequence of the Poisson process, the longitudinal
distances $\Delta$ between consecutive equipped are exponentially
distributed with the probability density
\begin{equation}
  f(\Delta)=\lambda e^{-\lambda \Delta},
\end{equation}
which is very well supported by empirical data \citep{IVC-Goldrain06}.

Particularly, given the full density $\lambda_{\rm tr}$ of equipped
cars on all lanes for the opposite driving direction, we may calculate
the cumulative probability distribution of the time $\tau$, after
which the message is available at the distance $r_{\rm u}$ upstream
from the position of message generation:
\begin{equation}
  P(\tau<t)=\Theta\left(t-\frac{r_{\rm u}-2R}{v_{\rm tr}}\right)
     \left( 1-e^{-\lambda_{\rm tr} \left(2 R+ v_{\rm tr}t-r_{\rm u} \right)}
     \right).\label{glg:transhop}
\end{equation}
Here, $R$ denotes the broadcast range of a sender/receiver unit,
$v_{\rm tr}$ is the (average) velocity on the opposite lanes, and the
Heavyside function $\Theta(x)$ is defined by $\Theta(x)=1$ for $x>0$
and $\Theta(x)=0$ otherwise.

\begin{figure}[th!]
  \begin{center}
  \includegraphics[width=73mm]{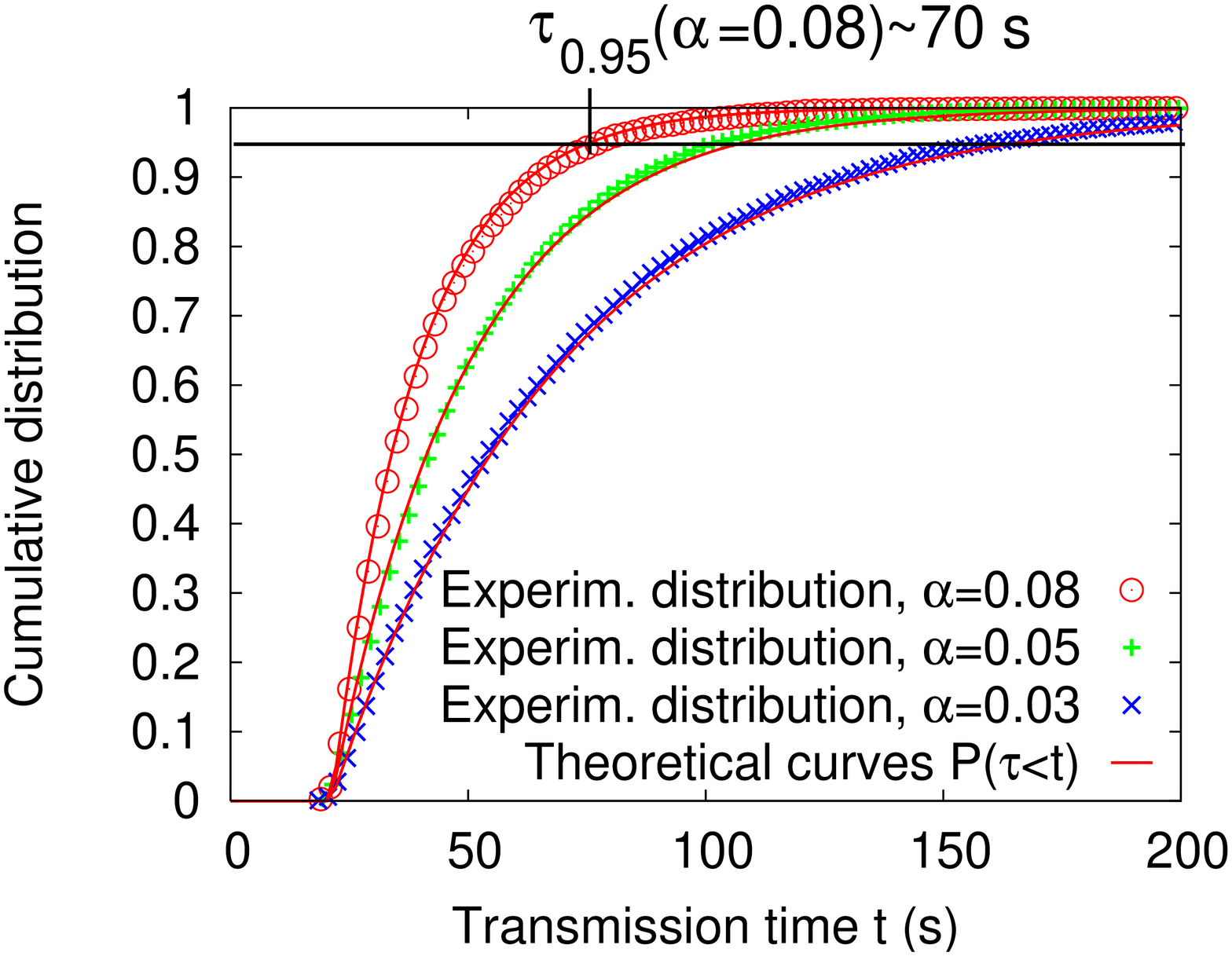}
  \includegraphics[width=73mm]{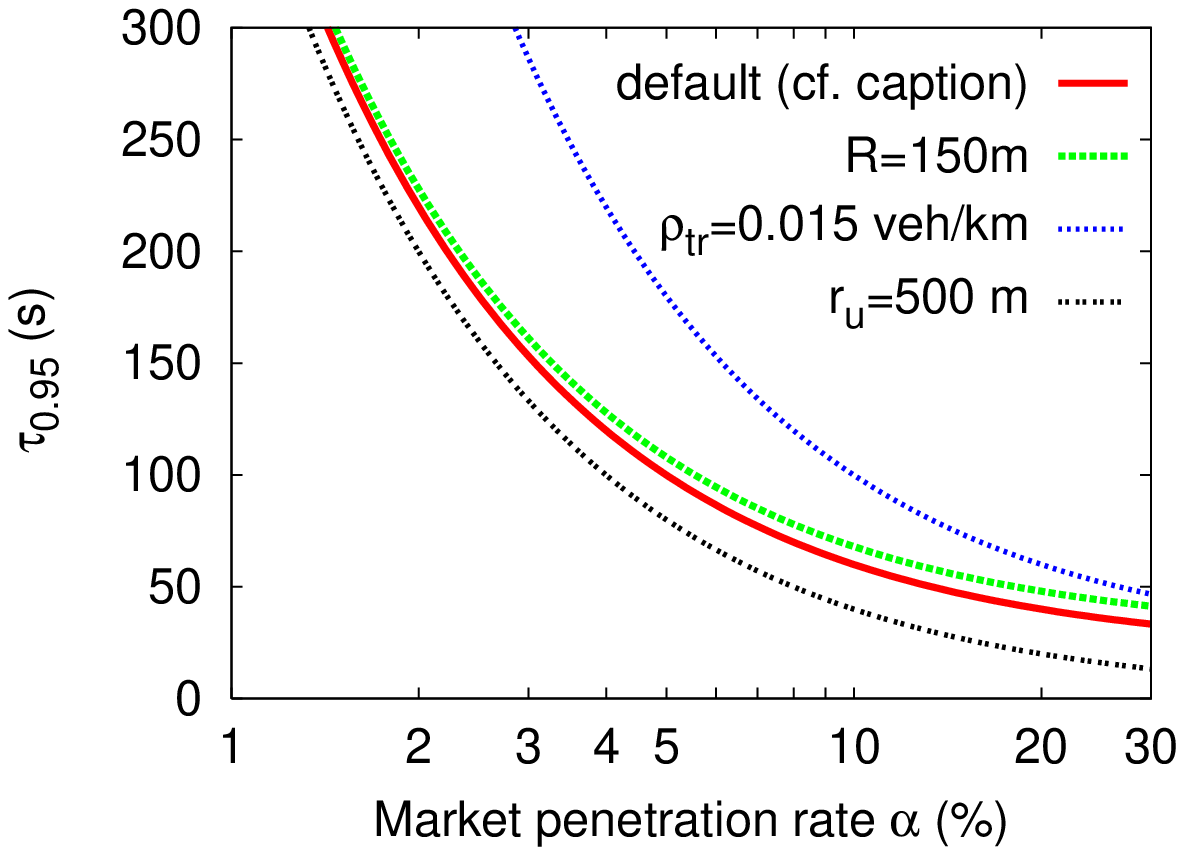}
  \end{center} \caption[]{Left diagram: Cumulative distribution of
  transmission times, if messages are transmitted via cars in the
  opposite driving direction.  Symbols correspond to simulation
  results and solid curves to the analytical result of
  Eq.~(\ref{heureka}). The assumed IVC parameters are the broadcast
  range $r_{\rm max}=250\,$m and the minimal delivery range $r_{\rm
  u}=1000\,$m. A chosen moderate inflow of $Q=1240$/h/lane resulted in
  a transmitter-vehicle velocity of $v_{\rm tr}=85\,$km/h and an
  overall traffic density of $\rho=29$/km in each direction. The
  simulations have been carried out with equipment rates of
  $\alpha=3$\%, $\alpha=5$\%, and $\alpha=8$\%.  With rising market
  penetration rate $\alpha$ the message transport becomes faster. The
  time $\tau_{0.95}$ indicates when 95\% of the messages are available
  $r_{\rm u}=1000\,$m upstream.\\ Right diagram: Investigation, how
  the efficiency of the message propagation in terms of $\tau_{0.95}$
  depends on the parameters by evaluating Eq.~\ref{heureka}.  The
  default parameters are $r_{\rm u}=1000\,{\rm m}$, $r_{\rm
  max}=250\,{\rm m}$, $\rho=0.03/{\rm m}$ and $v_{\rm tr}=90\,{\rm
  km/h}$.  Each of the other curves result by the variation of only
  one parameter, which reveals its impact.}  \label{fig:transExplain}
\end{figure}

The left diagram in Fig.~\ref{fig:transExplain} shows the
distributions of $\tau$ for several values of $\alpha$.  The
microscopic simulation approach allows for a detailed modeling of the
message broadcast and receipt mechanisms of IVC equipped vehicles (cf.
Fig.~\ref{fig:screenshot}). For a description of the traffic
simulation setup cf. the beginning of Section \ref{subsec:traffic}.
To obtain the statistics of message propagation, the equipped vehicles
have generated a 'dummy' message while crossing the position $x=5\,$km
in a freeway stretch of $10\,$km length, and the cycle time for the
communication has been set to $0.5\,$s.  The results show a very good
agreement with the analytical calculations (Eq.~\ref{glg:transhop}).
Note that the lower limit for $\tau$,
\begin{equation}
\tau_{\rm min}=\frac{r_{\rm u}-2R}{v_{\rm tr}}\,,
\end{equation}
 is realized, if the transmitter has an optimal position at
$t=0$, cf. Fig.~\ref{fig:sketch}.
%
%

Apart from this lower limit, which depends on $R$, $r_{\rm u}$, and
$v_{\rm tr}$, the value of $P(\tau<t)$ is strongly influenced by the
density of equipped vehicles in the opposite driving direction. This
becomes obvious, by looking at the expectation value
$\avg{\tau}$:
\begin{equation}
\avg{\tau}=\frac{r_{\rm u}-2R}{v_{\rm tr}}+\frac{1}{\lambda_{\rm
tr} v_{\rm tr}}.
\end{equation}
The higher the market penetration level $\alpha$, the higher are the
values for $ \lambda_{\rm tr}=\alpha\rho_{\rm tr}$, and the earlier a
message arrives upstream due to the decreasing second part of the
expression for $\avg{\tau}$ (the average waiting time for a
transmitter car).

Let us now consider the $95$th percentile of the IVC transmission time
as a quantity for the IVC efficiency. The definition 
\begin{equation}
  P(\tau \le \tau_{0.95})=0.95
\end{equation}
leads to the result
\begin{equation}\label{heureka}
  \tau_{0.95}= \frac{r_{\rm u}-2R}{v_{\rm tr}}
              +\frac{\ln(\frac{1}{0.05})}{\lambda_{\rm tr}v_{\rm tr}}.
 \end{equation}
This quantity is shown in Fig.~\ref{fig:transExplain} (right) for four
different scenarios. For a low equipment rate $\alpha$, it depends only
weakly on the broadcast range $R$ and the minimal propagation distance
$r_{\rm u}$, because the second term in Eq.~(\ref{heureka})
dominates: The traffic density $\rho_{\rm tr}$ on the
opposite lanes has a high impact.

Notice that in the limit of the upstream distance $r_{\rm u}\to
\infty$, the mean IVC propagation velocity $\avg{v}=\avg{\frac{r_{\rm
u}}{\tau}}$ converges to $v_{\rm tr}$, i.e., to the average speed of
the vehicles. So, at least in the long term, messages can propagate
faster via the transversal hopping mechanism than any congestion front
because the maximum upstream propagation of jam fronts is much smaller
than $v_\text{tr}$ \citep{BertMalik2004}.

\section{\label{sec:result}Simulation of jam-front detection and prediction}
%
\subsection{\label{subsec:traffic}Traffic simulation scenario}
For matters of illustration, we apply the proposed jam-front detection
(Sec.~\ref{sec:fronts}) and the message propagation via IVC
(Sec.~\ref{sec:messageProp}) to a specific traffic scenario.

As simulation scenario, we consider a homogeneous freeway section of
length $5\,$km with two independent driving directions and altogether
four lanes. The longitudinal movement of the vehicles are described by
the {\em Intelligent Driver Model} (IDM) \citep{Opus}, which is a
simple and realistic car-following model. The lane-changing decision
are based on the recently proposed model MOBIL \citep{MOBIL-TRB07}. We
have introduced heterogeneity by distributing the desired velocities
of the 'vehicle-driver units' according to a Gaussian distribution
with a standard deviation of $18\,$km/h around a mean speed of
$v_0=120$\,km/h. The other parameter values have been chosen according
to \cite{TGF05-ACC}. Notice that the details of the traffic
model do not influence the dynamics of message propagation via IVC,
which is our main focus here.

A given fraction $\alpha$ of the cars is randomly chosen to be
equipped with an IVC module. In the simulation, these vehicles
determine additionally their position by a 'satellite positioning
system' and feed their 'jam front detection device' by their own
velocity time series according to Sec.~\ref{sec:fronts}.
 
In one driving direction, we have triggered a stop-and-go wave
('Moving Localized Cluster', \citep{Phase,Martin-empStates}), while
traffic is free in the other driving direction. The inflow at both
upstream boundaries of the simulated freeway stretch is set to
$1800\,$vehicles/h/lane. Notice that the outflow (discharge rate) at
the downstream jam front is of the same order, so that the upstream
and the downstream jam front propagate with the characteristic speed
of about $15\,$km/h in upstream direction through the system.

\begin{figure}[th!]
  \begin{center}
    \includegraphics[width=12cm]{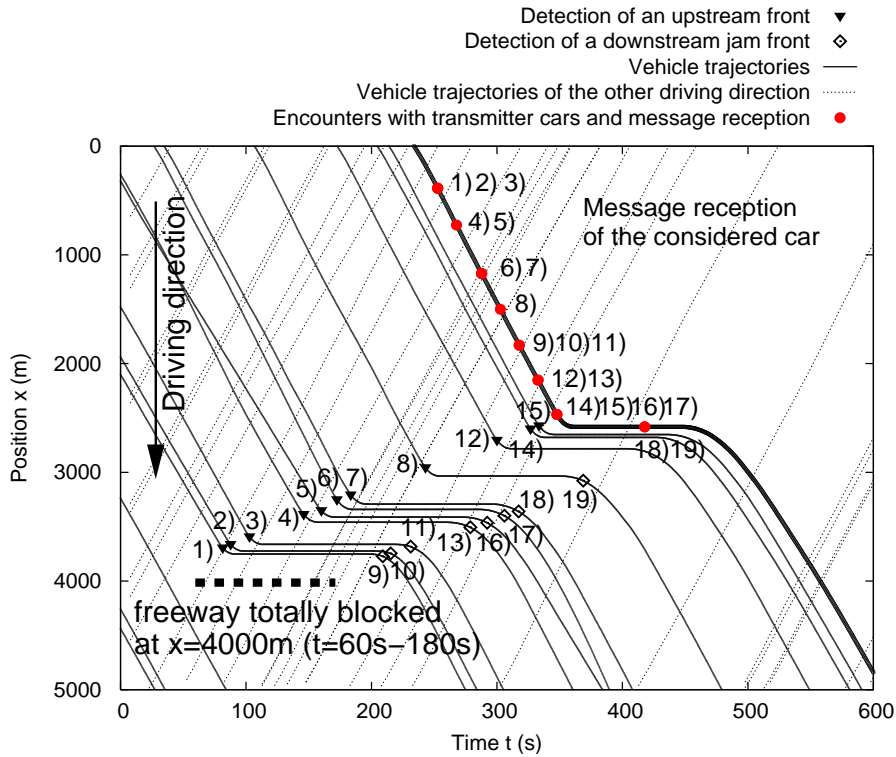}
\end{center}

 \caption{\label{fig:screenshot}Space-time diagram of the considered
 traffic scenario. The trajectories of IVC-equipped vehicles (only 3\%
 of all vehicles) are displayed by solid and dotted lines depending on
 the driving direction. The vehicles in the other driving direction
 serve as transmitter cars for the transversal hopping mechanism.  A
 temporary road blockage triggers a stop-and-go wave indicated by
 horizontal trajectory curves in one driving direction.  When cars
 encounter the propagating 'moving localized cluster', they broadcast
 messages about the detected position and time of the upstream jam
 front and the following downstream jam front. These message
 generations are represented by numbers. The reception of these
 messages by the considered vehicle (thick solid line) is indicated by
 the same numbers. The trajectories of the subsequent vehicles of this
 considered car are not shown. Notice that the crossing trajectories
 of equipped vehicles in the upper-left corner in the diagram refer to
 a passing maneuver due to different desired velocities}
\end{figure}

We consider an IVC equipment rate of $\alpha=3$\%. The resulting
trajectories and the sending and receiving events via transversal
hopping are illustrated in Fig.~\ref{fig:screenshot}. As already
pointed out in Sec.~\ref{IVC}, the distance of the equipped vehicles
often exceeds the broadcast range of $R=250\,$m, even in the region of
congested traffic. As shown in Fig.~\ref{fig:screenshot}, the
considered vehicle receives the first message about the upcoming
traffic congestion already $2\,$km before encountering the traffic
jam. Further received messages from other equipped vehicles are used
to confirm and update the predicted downstream traffic situation. 
%
%
%
\subsection{Interpretation of the received messages: Jam front prediction algorithm}
After the generation and propagation of traffic-related messages, the
received messages are finally used for a vehicle-based reconstruction
and prediction of the downstream traffic situation.
Each car in the microscopic simulation sorts the incoming messages
according to the reported jam-front type for a separate evaluation.
Within each message group, all messages that are not older than 120~s
(compared to the most recently received message) are considered for
the prediction: If this selection process yields two or more relevant
messages, the prediction for the jam front is based on linear
regression in the space-time plane, as the assumption of a constant jam
front velocity is well justified for the time scale of several
minutes, cf.~Sec.~2.3. In case of only one valid message, the reported
position in this message is regarded as the prediction of the jam
front position.

\begin{figure}[th!]
  \begin{center}
  \includegraphics[width=10cm]{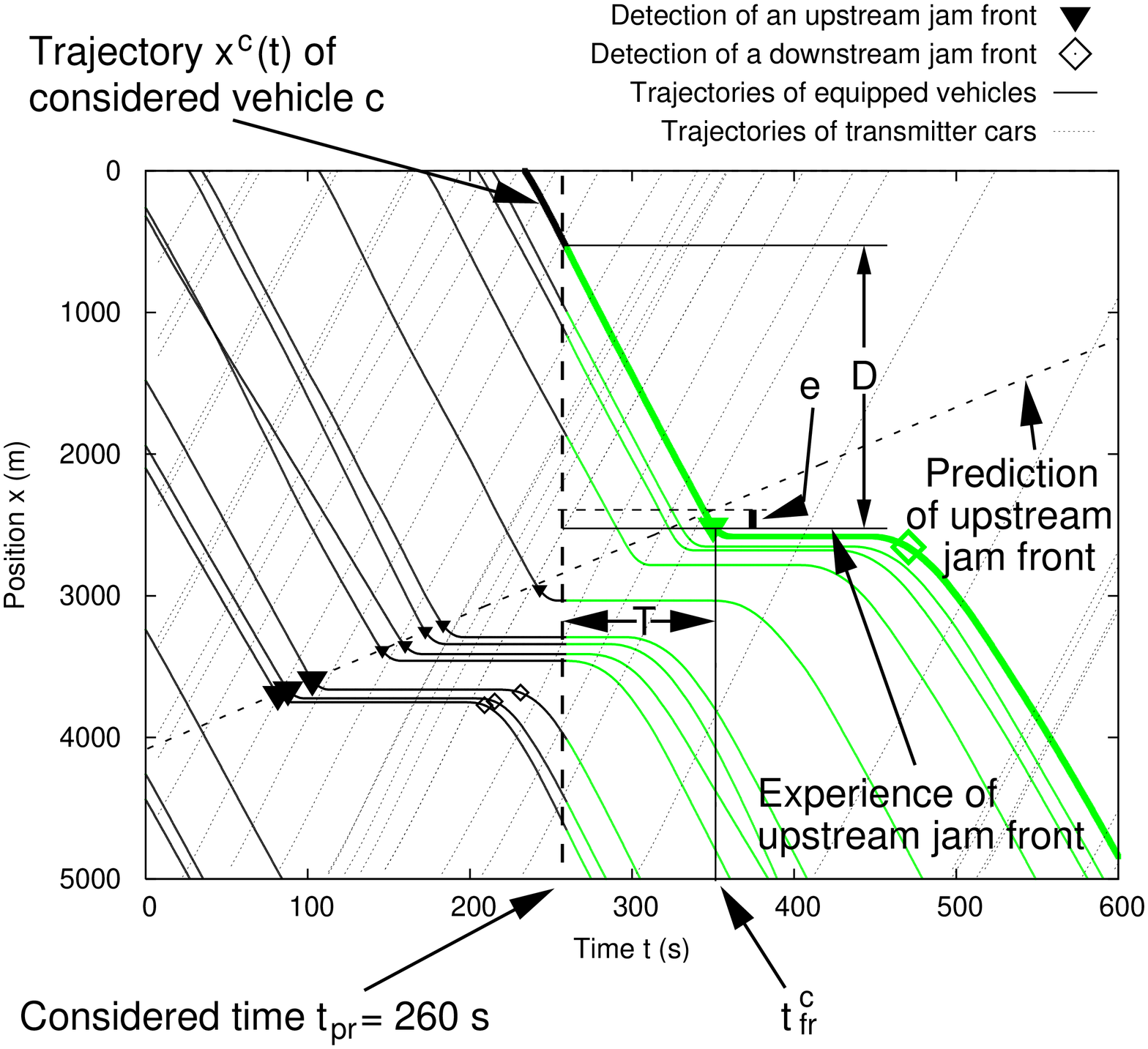}
  \includegraphics[width=10cm]{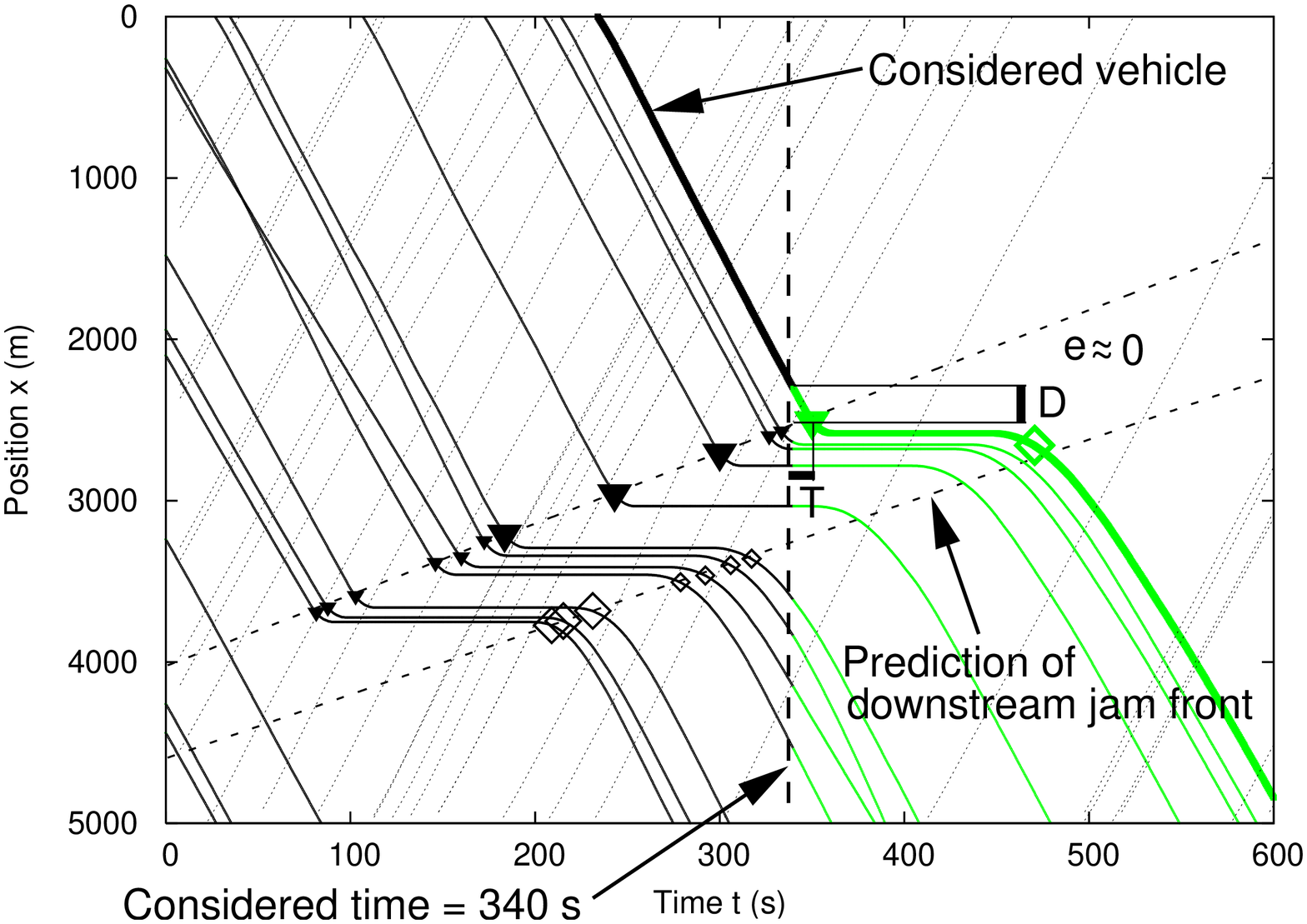}
\end{center}

\caption{\label{pmeasure}Two subsequent snapshots of our congestion
  front prediction as the considered vehicle (thick solid line)
  approaches the jam front. The symbols denote the generation of
  messages when a vehicle detects a jam front. Messages indicated by
  large symbols are used for the actual jam front prediction by the
  considered car, while small symbols correspond to outdated messages
  or messages that have not yet reached the considered car via IVC. The
  prediction quality is measured by the error $e$, which is defined by
  the difference between the real jam-front position as it will be
  detected by the considered car in the future and the actually
  predicted position for this event at a given time. Notice that the
  error $e$ can only be calculated {\it a posteriori} as it compares
  prediction and realization.}
\end{figure}

In order to analyze the quality of the jam-front prediction, we define
an error measure that compares the {\it prediction} calculated at a time
$t_{\text pr}$ with the actual {\it realization} of the jam front. Notice that
this error can only be calculated {\it a posteriori} as the considered
vehicle will pass the {\it real} jam-front at some time in the future.
As the forecast is carried out autonomously by each
equipped vehicle, we now consider a single car $c$ with its trajectory
$x^c(t)$. At a considered time $t_\text{pr}<t$, the car is located at
$x^c(t_\text{pr})$ and predicts the position of the jam front at time
$t$ by the linear regression function
$X_\text{fr}^{c\,,\,t_\text{pr}}(t)$. The prediction quality is
evaluated {\it ex post} when the vehicle $c$ {\it detects} itself the
jam front at a time $t_\text{fr}^c$ and at a position
$x^c(t_\text{fr}^c)$. Thus, the difference between the real jam-front
position $x^c(t_\text{fr}^c)$ and the predicted one for the time
$t_\text{fr}^c$ defines the error $e$ by
\begin{equation}\label{eq:predictError}
  e^{c}(t_\text{pr})= x^c (t_\text{fr}^{c}) -
  X_\text{fr}^{c\,,\,t_\text{pr}}\left( t_\text{fr}^{c}\right).
\end{equation}
Figure~\ref{pmeasure} illustrates the error $e$ for two snapshots at
the prediction times $t_\text{pr}=260\,$s and $340\,$s, respectively.
As shown in the diagrams, $e$ decreases when approaching the jam-front
as measured by the distance $D$.

%
%
Each equipped car predicts the jam fronts every two seconds according
to the communication cycle described in Sec.~\ref{IVC}. If no
messages arrive in one update cycle, no new prediction will be
estimated. After the simulation run, we determine the prediction error
$e$ 'off-line'. The prediction error $e$~(Eq.~\ref{eq:predictError})
depends on the time $t_\text{pr}$ and should, in average, decreases
when the considered car approaches the predicted jam front due to the
reception of new messages allowing for a better forecast. By a simple
time shift transformation,
\begin{equation}
T:=t_\text{fr}^c - t_\text{pr},
\end{equation}
$e$ depends on the time interval $T$ between the prediction time and
the time when the traffic jam front is reached.

\begin{figure}[th]
  \begin{center}
  \includegraphics[width=15cm]{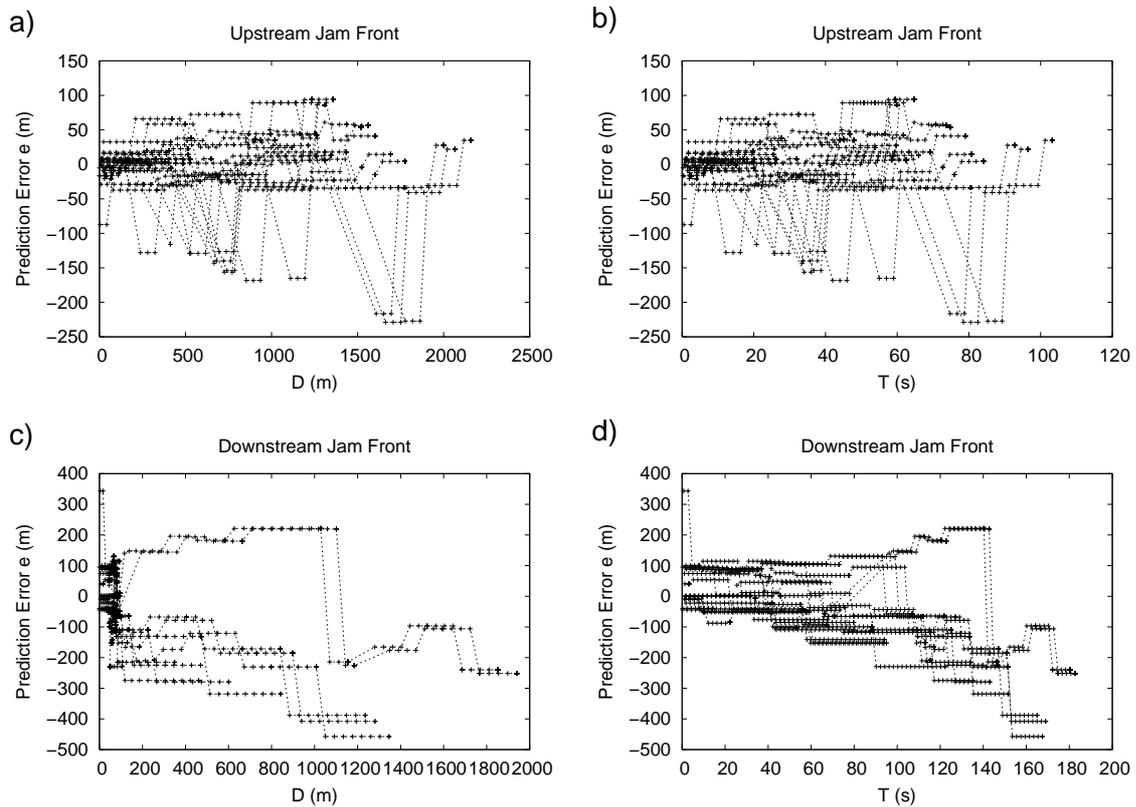}
\end{center}

 \caption{\label{evalues}The prediction errors $e$ of several equipped
 cars for the upstream jam front (a,b) and the downstream jam front
 (c,d) for the traffic scenario shown in
 Fig.~\protect\ref{pmeasure}. In order to guide the eyes, subsequent
 predictions of an equipped car are connected by a dotted line. The
 prediction (and, thus, the prediction error) is updated when a car
 gets new messages via inter-vehicle communication. The diagrams show
 the errors as a function of the spatial distance $D$ to the real jam
 front position in the future (a,c) and as a function of the remaining
 time $T$ (b,d). The errors decrease with decreasing $D$ and $T$ as
 the vehicles approach the jam fronts.  For $e>0$, the predicted jam
 front position is further upstream than the real jam front,
 cf. Fig.~\protect\ref{pmeasure}.}
\end{figure}

In Fig.~\ref{evalues} all single predictions in each car for a
scenario of Fig.~\ref{pmeasure} are shown. For matters of
illustration, the subsequent predictions of a single car are connected
by dotted lines.  In Fig.~\ref{evalues}\protect(a,b), it is shown, how the
error $e$ for the upstream jam front prediction depends on $D$ and
$T$.  In the case $D\to 0$ corresponding to $T\to 0$, the final
deviations $e$ of the considered vehicles do not exceed an interval of
about $\pm 50\,$m. The two diagrams show similar characteristics
because the velocity of the cars does not vary significantly so that
$D$ is essentially proportional to $T$.

Notice that this does not hold anymore when approaching a downstream
jam front as shown in Fig.~\ref{evalues}(c,d). A considered car spends
approximately $120\,$s in the jam, and, as a consequence, is already
close to the downstream jam front. Therefore, the car receives several
messages and updates predictions for $D\approx 50\,$m. This explains
the clustering of the symbols in Fig.~\ref{evalues}(c), while this
effect is not relevant in the representation of Fig.~\ref{evalues}(d).
In the latter diagram, it can be seen, that for a car in the
congestion zone ($T<120$~s) the mean frequency of prediction changes
is decreased compared to the situation in the free flow state because
a non-moving vehicle encounters less frequently transmitter cars of
the other driving direction and thus receives less new messages per
time than a moving vehicle. For $T\to 0$, the errors for predicting
the downstream jam front are restricted to values of the order of $\pm
100\,$m.  It turned out, that the reason is given by the
characteristics of the downstream jam front.  As visible in
Fig.~\ref{pmeasure} and Fig.~\ref{fig:screenshot}, the downstream jam
front becomes smoother after some time -- the cars spend more time
(and space) in the acceleration process. This is due to the traffic
dynamics in this special situation. The first car leaving the jam has
no leader car in front, and thus its acceleration is higher compared
to subsequent cars leaving the jam.  This affects the detection of the
front, which gets more and more delayed compared to the first
detection processes at $t=200$~s.  Thus the prediction of the front
position is likely to be further upstream than the actual position
detected according to Eq.~(4), which leads to prediction errors of
$e>0$ according to applied definition of $e$.  The second clustering
of $e$ in the interval between 50 and 100~m seems to be a result of
this spatiotemporal curvature of the jam front.  In summary, it has
to be stated, that the prediction errors are also caused by
uncertainties in determining the actual position of the downstream
front by Eq.~(4).

\clearpage
\section{\label{sec:dis}Summary, critical discussion, and outlook}

In this contribution, a completely distributed freeway traffic
information system based on inter-vehicle communication (IVC) has been
introduced.  It involves an autonomous, vehicle-based jam front
detection, the information transmission via IVC, and the prediction of
the spatial position of jam fronts by reconstructing the
spatiotemporal traffic situation based on the transmitted information.
The whole system is simulated within an integrated microscopic traffic
simulation framework. The function of its communication module has
been explicitly validated by comparing the simulation results with
analytical calculations for the message propagation. By means of
simulations, we have shown that the algorithms for a congestion-front
recognition, message transmission, and processing predict reliably the
existence and position of jam fronts for vehicle equipment rates as
low as 3\%. The prediction error decreases when a car is approaching
the front, but does not go to zero due to uncertainties in determining
the exact positions of a jam front. Notice that a reliable mode of
operation for small market penetrations is crucial for the successful
introduction of the technology proposed in this paper. The obtained
accurate prediction results can be used as (non-local) input for a
traffic-adaptive ACC system that changes its driving
characteristics according to the local traffic situation as recently
proposed by \cite{ACC-Arne-TRB07}. Furthermore, the timely knowledge
about the upcoming traffic situation on a scale of minutes and a few
kilometers offers new possibilities for vehicle-based advanced driver
information systems.

Our feasibility study has been based on the following main assumptions
and mechanisms: (i) Inter-vehicle communication allows for a reliable
exchange of small data packages between vehicles in different driving
directions with speed difference of up to $300\,$km/h. The broadcast
range is limited, e.g., to $250\,$m.  (ii) Traffic in the opposite
driving direction is free. In the traffic simulations, stationary
traffic flow conditions have been assumed, but this is not a
precondition for the functionality of the proposed concept.  (iii) The
propagation speed of the dynamic jam fronts changes little on time
scales of several minutes.
 
A few short remarks concerning the presented approach may summarize
the limitations of our study. The currently available communication
technology, and the future standards of wireless data transmission
between vehicles, allow for a fast and reliable dissemination of small
traffic-related messages (as needed for our concept) even for a high
density of equipped vehicles. However, it should be mentioned that
other applications may use the provided bandwidth as well. We suppose
that there will be a balanced distribution of the resources needed for
traffic safety, traffic information, and other applications.

For a small market penetration of IVC-equipped vehicles, the
characteristics and efficiency of the message transport rely to a
large extend on the density of equipped vehicles of the opposite
driving direction and their driving speed. Note, that in rush hours,
often only one of the driving directions is congested. The same
applies for traffic congestion caused by accidents. Thus in most of
the cases, free traffic flow of the opposite driving direction can be
assumed. With respect to the considered application in
'traffic-adaptive' ACC systems, it is essential to predict traffic-jam
fronts. Within short time scales of 5 to 10 minutes, the velocity of
the traffic-jam fronts can be assumed to be constant in the most
cases, in particular for downstream jam fronts.  Further research is
necessary in order to improve the proposed prediction model and to
reduce the prediction error. In particular, this applies to traffic
situations with several traffic jam fronts of the same type and to
special situations, in which the jam-front velocity changes, e.g.,
because of the clearance after an accident.

\section*{Acknowledgments:}
The authors would like to thank Hans-J\"urgen Stauss for the excellent
collaboration and the Volkswagen AG for partial financial support
within the BMBF project INVENT.



\end{document}